# Improved Stability and Controller Design Criteria for Two-dimensional Differential-Algebraic-Equation Systems via LMI Approach

1st Abdolah RoshanaeeDeh
M.Sc. Graduate
Department of Control Engineering
Amirkabir university of technology
Tehran, Iran
roshanaeedeh@aut.ac.ir

2nd Hajar Atrianfar
Associate Professor
Department of Control Engineering
Amirkabir university of technology
Tehran, Iran
atrianfar@aut.ac.ir

3rd Masoud Shafiee
Full Professor
Department of Control Engineering
Amirkabir university of technology
Tehran, Iran
mshafiee@aut.ac.ir

*Abstract*—This paper deals with issues related to the asymptotic stability testing and the achievement of controllers for the two-dimensional Rosser model in Differential-Algebraic-Equations systems (DAEs). Sufficient stability criteria, concerning the Lyapunov approach, are presented using a collection of Linear-Matrix-Inequalities (LMIs) for two-dimensional DAEs. Subsequently, we derive a collection of sufficient conditions to determine the presence of both state- and output-feedback controllers. Our proposed methods have obviated the need for decomposing the two-dimensional DAEs into two algebraic and differential parts. Ultimately, an industrial example illustrates the efficiency of the suggested methodologies.

## I. INTRODUCTION

### A. Background and Motivation

**P**ERPETUALLY , one cannot be certain that the description of physical systems in the one-dimensional state-space representation (ordinary differential equations) without algebraic constraints provides a comprehensive and unblemished description of all physical and industrial systems. Frequently, in the modeling of physical systems, the requirement for accurate system representation, especially in contexts demanding high accuracy, such as robotics [1], biological systems [2], [3], and similar industrial scenarios, necessitates multi-dimensional modeling. This entails considering the system's dependency on multiple independent variables, making it an essential aspect of state-space modeling [4]. Therefore, it is necessary to generalize the state-space differential equations of the system. This crucial concept has given rise to an innovative approach in modeling state-space systems, recognized as "Differential-Algebraic-Equation system" (DAEs).

DAEs, also known as singular, descriptor, implicit, or generalized state-space systems, have garnered significant interest in recent decades [5], [6]. This system modeling approach, pioneered by Luenberger [7], constitutes a significant category of dynamic system models, holding importance from both practical and theoretical perspectives. Its importance stems from the ability to depict the configuration of physical systems while incorporating algebraic relationships among state variables of system [8]. The system index represents the degree of interdependence among the state variables within the state-space model of DAEs [9]. Additionally, the number of independent variables to which the state variables of the state-space model are linked is referred to as the dimension of the DAEs [10].

DAEs can exist in both one-dimensional and multi-dimensional forms. In other words, their dynamic equations encompass partial derivatives concerning multiple independent variables, reflecting the system's dependence on these variables [10]. The complexity introduced by expanding dimensions in DAEs increases the intricacy involved in stability analysis and control. This complexity is especially noticeable in comparison with state-space systems described by ordinary differential equations. Moreover, these systems pose relatively greater challenges in terms of control objectives. This is because control theories suitable for conventional state-space systems do not seamlessly extend to these specific systems.

The attainment of desired control objectives and the design of controllers by creating a collection of Linear-Matrix-Inequalities (LMIs) have consistently remained the central topics of the researches [11], [12].

### B. Research Objectives and Contributions

The framework of this research is outlined in more details. In Section II, The two-dimensional DAEs with the Rosser model and its admissibility conditions are introduced. In Section III, a collection of innovative sufficient criteria, presented as LMIs, is provided for stability testing. Subsequently, novel criteria, represented as LMIs, are established for attaining state- and output-feedback controllers. Eventually, in Section IV, the implementation from the suggested methods on the heat transfer equa-tion is addressed, the efficiency of these recommended approaches is demonstrated.

In [8], [19], [20], a prevalent technique for stability analysis involves decomposing DAEs into differential and algebraic subsystems. However, in this article, a novel method has been introduced that obviates the necessity of such decomposition, while also demonstrating significantly faster operation compared to their methodologies.

The innovations presented in this paper are expressed as follows:

- Unlike previous works such as [8], [13]–[17], which derived stability criteria and achieved various controllers exclusively for two-dimensional systems in the form of LMIs, our proposed methods are also applicable and extendable to two- and n-dimensional DAEs.
- A common approach has been utilized by recent researchers [8], [14]–[17], [20] for stability testing or designing controller in two-dimensional DAEs involves decomposing the system into two algebraic and differential parts. However, this method becomes complex for high-dimensional and high-order systems. Our approach obviates the need for system decomposition into differential and algebraic parts.

## II. PRELIMINARIES

In this paper, the objective is to avoid the need for decomposing two-dimensional DAEs into two algebraic and differential equations for stability testing and designing controllers, both for state-feedback and output-feedback. To achieve this, we provide the necessary prerequisites and definitions.

At first, we should note that discrete-time state-space models with ordinary difference equations cannot model all systems. Therefore, a more comprehensive state-space modeling technique for discrete-time two-dimensional DAEs, known as the Rosser model, is presented as follows:

$$E \begin{bmatrix} x^h(i+1,j) \\ x^v(i,j+1) \end{bmatrix} = A \begin{bmatrix} x^h(i,j) \\ x^v(i,j) \end{bmatrix} \quad (1)$$

In this context, $x^h \in \mathbb{R}^{n_h}$ is defined as a state vector representing the horizontal components, $x^v \in \mathbb{R}^{n_v}$ is defined as a state vector representing the vertical components, and $A \in \mathbb{R}^{n \times n}$, where $n = n_v + n_h$, as the coefficient matrix displaying the system's state variables. Since $E \in \mathbb{R}^{n \times n}$ is a singular matrix, with a rank reduction, it is rank$(E) = n_1 \leq n$. Following an initial introduction to the stability and admissibility of two-dimensional DAEs, a more detailed explanation will be presented.

***Definition 1:*** [10]

- The system (1) is classified as regular if $\det [EI(z_1, z_2) - A] \neq 0$, satisfying the condition $I(z_1, z_2) = \text{diag} \{z_1 I_{n_1}, z_2 I_{n_2}\}$.
- The system is known as causal when the degree of $\det (EI(z_1, z_2) - A)$ is equivalent to the rank of matrix E.

- The concept of stability is relevant to the system represented by (1) if and only if the set $\lambda(E, A) \subset D_{\text{int}}(0, 1)$ holds true.
- The system (1) is called admissible if it fulfills the criteria of being regular, causal, and stable.

Subsequently, lemmas that are used to prove the theorems in this article are presented.

***Lemma 1:*** (Congruence [18]) Two matrices are denoted as $S$ and $D$. If $S$ exhibits characteristics of being negative definite (or positive definite) and $D$ is a full-rank matrix, then the matrix $D^T S D$ is also negative definite (or positive definite).

***Lemma 2:*** (Schur complement [18]) With constant matrices $\Omega_1$, $\Omega_2$, and $\Omega_3$ of suitable dimensions, where $\Omega_1$ and $\Omega_3$ are symmetric, the conditions $\Omega_3 > 0$ and $\Omega_1 + \Omega_2^T \Omega_3^{-1} \Omega_2 < 0$ are satisfied if and only if:

$$\begin{bmatrix} \Omega_1 & \Omega_2^T \\ \Omega_2 & -\Omega_3 \end{bmatrix} < 0. \quad (2)$$

## III. MAIN RESULTS

This section begins by deriving stability conditions using LMIs. Following this, designing of state- and output-feedback controllers for two-dimensional DAEs is presented. The path of stability analysis and controller design follows a unique approach, claiming its superiority over existing methods in recent researches [8], [13]–[17], [21].

### A. Stability Analysis

The mentioned references decompose two-dimensional DAEs into two components: differential and algebraic equations. In the following, we present a method that, without the need for decomposing the system into separate differential and algebraic parts, tests the stability of two-dimensional DAEs in a unique approach. However, to define Lyapunov functions for two-dimensional DAEs in the following sections, an abstract comprehension of system decomposition is required. We proceed to explore this concept for further research.

Due to the rank of $(E) = n_1 \leqslant n$, there is always a potential to discover two invertible matrices, labeled as $U$ and $V$ in a way that:

$$UEV = \begin{bmatrix} I_{n_{h1}} & 0 & 0 & 0 \\ 0 & I_{n_{v1}} & 0 & 0 \\ \hline 0 & 0 & 0 & 0 \\ 0 & 0 & 0 & 0 \end{bmatrix}, UAV = \begin{bmatrix} A_1 & A_2 \\ \hline A_3 & A_4 \end{bmatrix}, \quad (3)$$

Consequently, just as the two-dimensional system is decomposed into two parts, the Lyapunov function is also split into two dimensions, and its differences are computed concerning the two independent variables $i, j$. The Lyapunov candidate function for two-dimensional DAEs is considered in the form of (4).

$$V \left( \begin{bmatrix} x^h(i,j) \\ x^v(i,j) \end{bmatrix} \right) = V^h \left( x^h(i,j) \right) + V^v \left( x^v(i,j) \right) \quad (4)$$

where
$$\begin{cases} V^h\left(x^h(i,j)\right) = \left(E^h x^h(i,j)\right)^T P^h E^h x^h(i,j) \\ V^v\left(x^v(i,j)\right) = \left(E^v x^v(i,j)\right)^T P^v E^v x^v(i,j) \end{cases} \quad (5)$$

*Theorem 1:* The two-dimensional DAEs, as represented by (1), achieve asymptotic stability when a symmetric matrix $P = \text{diag}(P_h, P_v)$ can be found such that the following LMIs hold:
$$\begin{bmatrix} -E^T PE & A^T P \\ * & P \end{bmatrix} < 0 \quad (6)$$
$$E^T PE \geq 0$$

*Proof 1:* By taking the derivative of both sides of (4), we can derive the following result:
$$\begin{aligned} \Delta V\left(\begin{bmatrix} x^h(i,j) \\ x^v(i,j) \end{bmatrix}\right) &= \Delta V^h\left(x^h(i,j)\right) + \Delta V^v\left(x^v(i,j)\right) \\ &= V^h\left(x^h(i+1,j)\right) - V^h\left(x^h(i,j)\right) \\ &+ V^v\left(x^v(i,j+1)\right) - V^v\left(x^v(i,j)\right) \ . \end{aligned} \quad (7)$$

From (5) and (7), we can calculate the rate of change of the Lyapunov function:
$$\begin{aligned} \Delta V&\left(\begin{bmatrix} x^h(i,j) \\ x^v(i,j) \end{bmatrix}\right) \\ &= \left(E^h x^h(i+1,j)\right)^T P^h E^h x^h(i+1,j) \\ &- \left(E^h x^h(i,j)\right)^T P^h E^h x^h(i,j) \\ &+ \left(E^v x^v(i,j+1)\right)^T P^v E^v x^v(i,j+1) \\ &- \left(E^v x^v(i,j)\right)^T P^v E^v x^v(i,j) \end{aligned} \quad (8)$$

Accordingly, the initial constraint of the Lyapunov function, denoted by $E^T PE \geq 0$, must hold true. This condition serves as an initial prerequisite within the set of LMIs presented in (6).

Continuing further, under applying (1) to (8), we have:
$$\begin{aligned} \Delta V\left(\begin{bmatrix} x^h(i,j) \\ x^v(i,j) \end{bmatrix}\right) &= [x^{hT}(i,j)A^{hT}] P^h [A^h x^h(i,j)] \\ &- x^{hT}(i,j)E^{hT} P^h E^h x^h(i,j) \\ &[x^{vT}(i,j)A^{vT}] P^h [A^v x^v(i,j)] \\ &- x^{vT}(i,j)E^{vT} P^h E^v x^v(i,j) \end{aligned} \quad (9)$$

Now, considering the fact that coefficient matrices $E$ and $A$ possess the potential for recoupling, we will have:
$$\begin{aligned} \Delta V\left(\begin{bmatrix} x^h(i,j) \\ x^v(i,j) \end{bmatrix}\right) &= [x^T(i,j)A^T] P[Ax(i,j)] \\ &- x^T(i,j)E^T PE x(i,j) \end{aligned} \quad (10)$$

By factoring $x^T(i,j)$ from the left side and $x(i,j)$ from the right side, it results in:
$$\Delta V\left(\begin{bmatrix} x^h(i,j) \\ x^v(i,j) \end{bmatrix}\right) = x^T(i,j) \left[A^T PA - E^T PE\right] x(i,j) \quad (11)$$

On the other hand, it is known that the rate of change of the Lyapunov function must be negative. Therefore, the right side of the matrix in (11) needs to be evaluated as negative-definite. This leads us to formulate:
$$x^T(i,j) \underbrace{\left[A^T PA - E^T PE\right]}_{\Psi} x(i,j) < 0 \quad (12)$$

The vector of decision variables for problem (12) is defined as $\eta = x(i,j)$ and the LMI is reformulated as $\eta^T \Psi \eta < 0$.

Due to the presence of a matrix $E$ with rank deficiency in $\Psi$, the possibility of the problem (12) being infeasible is not insignificant. To amend the common stability criteria (12), leveraging the lemma 2, we have:
$$\begin{bmatrix} -E^T PE & A^T \\ * & P^{-1} \end{bmatrix} < 0 \quad (13)$$

The inequality (13) is infeasible due to the presence of $P^{-1}$. Thus, by using lemma 1, we reach the final LMI that defines the conditions of the theorem's scenario. The proof for Theorem (1) is now concluded. This innovative process is also applicable to designing controllers.

### B. State-feedback controller

To achieve and develop controllers on state-feedback structure, the system (1) in a closed-loop configuration is described as follows:
$$\begin{aligned} E\begin{bmatrix} x^h(i+1,j) \\ x^v(i,j+1) \end{bmatrix} &= (A + BK) \begin{bmatrix} x^h(i,j) \\ x^v(i,j) \end{bmatrix} \\ y(i,j) &= F \begin{bmatrix} x^h(i,j) \\ x^v(i,j) \end{bmatrix} \end{aligned} \quad (14)$$

where the input $u(i,j) = Kx(i,j)$ for controlling the system with coefficient matrix $B$ is incorporated into the system (1).

*Theorem 2:* Presume that the matrix $B$ has full row rank. The system (14), when subjected to the state-feedback controller $u = Kx(i,j)$, will achieve admissibility, internal stability, and impulse-free. This holds true if symmetric matrices, exemplified by $P = \text{diag}(P_h, P_v) \in \mathbb{R}^{(n_h+n_v)\times(n_h+n_v)}$ and $\tilde{Z} \in \mathbb{R}^{(n_h+n_v)\times(n_h+n_v)}$, exist and satisfy the following LMIs:
$$\begin{bmatrix} -E^T PE & A^T P + \tilde{Z} \\ * & P \end{bmatrix} < 0 \quad (15)$$
$$E^T PE \geq 0$$

where $\tilde{Z} = K^T B^T P$.

*Proof 2:* For this case, the Lyapunov function is defined as (4). When we apply the system in a closed-loop configuration, (14), in (8), it leads to:
$$\begin{aligned} \Delta V&\left(\begin{bmatrix} x^h(i,j) \\ x^v(i,j) \end{bmatrix}\right) = \\ &\left[x^{hT}(i,j)\left(A^h + B^h K^h\right)^T\right] P^h \left[\left(A^h + B^h K^h\right) x^h(i,j)\right] \\ &- x^{hT}(i,j)E^{hT} P^h E^h x^h(i,j) \\ &+ \left[x^{vT}(i,j)\left(A^v + B^v K^v\right)^T\right] P^v \left[\left(A^v + B^v K^v\right) x^v(i,j)\right] \\ &- x^{vT}(i,j)E^{vT} P^v E^v x^v(i,j) \end{aligned} \quad (16)$$

Recouping coefficient matrices $E$, $A$, $B$, and $P$ will result in:

$$\Delta V\left(\begin{bmatrix} x^h(i,j) \\ x^v(i,j) \end{bmatrix}\right) \\ = x^T(i,j)\left[(A+BK)^T P(A+BK) - E^T PE\right] x(i,j) \quad (17)$$

Following that, lemma 2 is applied, resulting in:

$$\begin{bmatrix} -E^T PE & A^T + K^T B^T \\ * & P^{-1} \end{bmatrix} < 0 \quad (18)$$

Now, by using the lemma 1, the problem of the existence of a non-linear matrix $P^{-1}$ can be solved and obtained:

$$\begin{bmatrix} -E^T PE & A^T P + K^T B^T P \\ * & P \end{bmatrix} < 0 \quad (19)$$

With the understanding that $K^T B^T P$ is nonlinear, a change of variable denoted as $\tilde{Z} = K^T B^T P$ results in the conversion of the matrix inequality (19) into linear forms. This transformation serves as the basis for proving Theorem (2). □

State-feedback controllers can prove highly advantageous in industrial and robotics applications. However, in situations where state variables are unavailable, our novel approach can still be applied to achieve and develop output-feedback controllers for DAEs.

*C. Output-feedback controller*

In this subsection, assuming that the state variables are not directly available in the system (1), the output-feedback controller is achieved through the following processes:

$$E\begin{bmatrix} x^h(i+1,j) \\ x^v(i,j+1) \end{bmatrix} = (A - BKF)\begin{bmatrix} x^h(i,j) \\ x^v(i,j) \end{bmatrix} \\ y(i,j) = F\begin{bmatrix} x^h(i,j) \\ x^v(i,j) \end{bmatrix} \quad (20)$$

***Theorem 3:*** Let's presume the matrix $B$ has full row rank. The solvability of the problem of controlling the system using output-feedback for two-dimensional DAEs with $u(i,j) = -Ky(i,j)$ can be achieved if symmetric matrix $P = \text{diag}(P_h, P_v)$ exists, where $P \in \mathbb{R}^{(n_h+n_v)\times(n_h+n_v)}$, and satisfies the following conditions:

$$\begin{bmatrix} -E^T PE & A^T P - F^T \tilde{X} \\ * & P \end{bmatrix} < 0 \\ E^T PE \geq 0 \quad (21)$$

where $\tilde{X} = K^T B^T P$.

***Proof 3:*** By applying (20) to (8), we will have:

$$\Delta V\left(\begin{bmatrix} x^h(i,j) \\ x^v(i,j) \end{bmatrix}\right) \\ = \left[x^{hT}(i,j)\left(A^h - B^h K^h F^h\right)^T\right] \\ P^h\left[(A^h - B^h K^h F^h) x^h(i,j)\right] \\ - x^{hT}(i,j) E^{hT} P^h E^h x^h(i,j) \\ + \left[x^{vT}(i,j)\left(A^v - B^v K^v F^v\right)^T\right] \\ P^v\left[(A^v - B^v K^v F^v) x^v(i,j)\right] \\ - x^{vT}(i,j) E^{vT} P^v E^v x^v(i,j) \quad (22)$$

Then, $\Delta V$ can be obtained:

$$x^T(i,j)\underbrace{\left[(A-BKF)^T P(A-BKF) - E^T PE\right]}_{\Psi} x(i,j) < 0 \quad (23)$$

The matrix inequality of (23) is infeasible. Using lemmas 1 and 2 leads to:

$$\begin{bmatrix} -E^T PE & A^T - F^T K^T B^T P \\ * & P \end{bmatrix} < 0 \quad (24)$$

By introducing a change of variable as $\tilde{X} = K^T B^T P$, Theorem (3) can be established. □

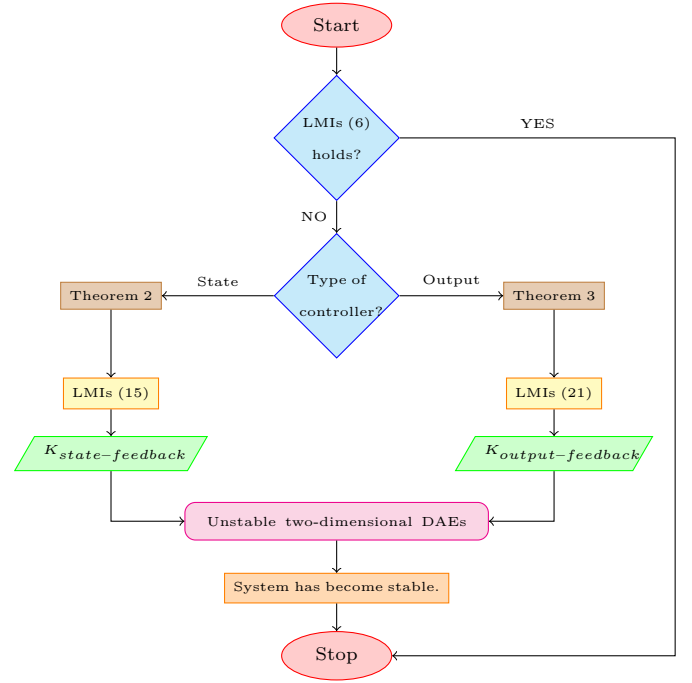

Fig. 1. A block diagram for stability checking and the design of controllers for two-dimensional DAEs.

Fig. 1. illustrates the systematic process of this paper. Using this graph, one can systematically organize the Theorems 1, 2, and 3 for two-dimensional DAEs.

This method can be extended for the design of various controllers such as dynamic, robust, adaptive, and so on, for two-dimensional DAEs. These controllers can be effectively utilized in industry and robotics to achieve control objectives.

## IV. CASE STUDY

In this part, an illustrative example is explored to showcase that existing heat transfer systems in the industry can be effectively modeled using two-dimensional DAEs.

As indicated in [4], the distribution of thermal energy in the heat transfer tubes of chemical reactors is represented by (25).

$$\frac{\partial T(x,t)}{\partial x} = -\frac{\partial T(x,t)}{\partial t} - T(x,t) + U(x,t) \quad (25)$$

In (25), $T(x,t)$ signifies the temperature at the designated position $x$ and time $t$, while $U$ denotes the operating temperature of the system.

Now, by discretizing the system utilizing the limited backward difference approximation method, the result is as follows:

$$\begin{aligned}\frac{\partial T(x,t)}{\partial x} &\simeq \frac{1}{\Delta x}[T(i\Delta x, j\Delta t) - T(i\Delta x - \Delta x, j\Delta t)] \\ \frac{\partial T(x,t)}{\partial t} &\simeq \frac{1}{\Delta t}[T(i\Delta x, j\Delta t) - T(i\Delta x, j\Delta t - \Delta t)]\end{aligned} \quad (26)$$

By inserting (26) into (25), the system (25) is discretized:

$$\begin{aligned}&\frac{1}{\Delta x}[T(i\Delta x, j\Delta t) - T(i\Delta x - \Delta x, j\Delta t)] = \\ &- \frac{1}{\Delta t}[T(i\Delta x, j\Delta t) - T(i\Delta x, j\Delta t - \Delta t)] \\ &- T(i\Delta x, j\Delta t) + U(i\Delta x, j\Delta t)\end{aligned} \quad (27)$$

To transform (27) into two-dimensional DAEs on Rosser state-space model (1), we introduce the following definitions for the state variables:

$$\begin{aligned}x^h(i,j) &= T(i\Delta x - \Delta x, j\Delta t) \\ x^v(i,j) &= T(i\Delta x, j\Delta t)\end{aligned} \quad (28)$$

Then, it results in:

$$\begin{aligned}&\begin{bmatrix} 1 & 0 \\ 0 & 0 \end{bmatrix}\begin{bmatrix} x^h(i+1,j) \\ x^v(i,j+1) \end{bmatrix} = \begin{bmatrix} \frac{\Delta t}{\Delta x + \Delta t + \Delta x \Delta t} & 0 \\ -\frac{\Delta t}{\Delta x + \Delta t + \Delta x \Delta t} & 1 \end{bmatrix}\begin{bmatrix} x^h(i,j) \\ x^v(i,j) \end{bmatrix} \\ &+ \begin{bmatrix} \frac{\Delta x \Delta t}{\Delta x + \Delta t + \Delta x \Delta t} \\ -\frac{\Delta x \Delta t}{\Delta x + \Delta t + \Delta x \Delta t} \end{bmatrix} u(i,j)\end{aligned} \quad (29)$$

It is assumed that $\Delta x = \Delta t = 0.1$ holds, the Rosser model of system (25) will finally be obtained as follows:

$$\begin{aligned}&\begin{bmatrix} 1 & 0 \\ 0 & 0 \end{bmatrix}\begin{bmatrix} x^h(i+1,j) \\ x^v(i,j+1) \end{bmatrix} = \begin{bmatrix} 0.476 & 0 \\ -0.476 & 1 \end{bmatrix}\begin{bmatrix} x^h(i,j) \\ x^v(i,j) \end{bmatrix} \\ &+ \begin{bmatrix} 0.047 \\ -0.047 \end{bmatrix} u(i,j) \\ &y(i,j) = \begin{bmatrix} 0.1 & 0.1 \end{bmatrix}\begin{bmatrix} x^h(i,j) \\ x^v(i,j) \end{bmatrix}\end{aligned} \quad (30)$$

The stability testing and designing controllers for industrial system (30) can be explored by utilizing Theorems 1, 2, and 3.

To perform a stability test for the system (30) under the initial conditions $x_h(0,0) = (i_0, j_0)$ and $x_v(0,0) = (i_0, j_0)$, we apply Theorem 1, and upon observing that the LMIs are infeasible, it is concluded that the system (30) is unstable.

MATLAB simulations of the system described by (30) are used to create a comprehensive illustration of the open-loop system's dynamic behavior in Fig. 2.

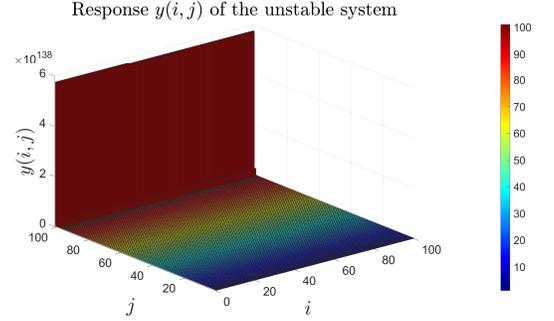

Fig. 2. Zero input response of the system (30).

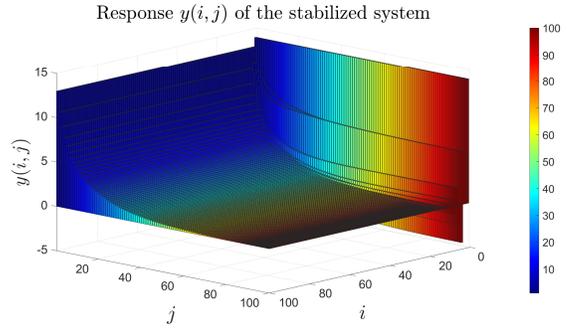

Fig. 3. The output of system (30) after stabilization by state-feedback controller.

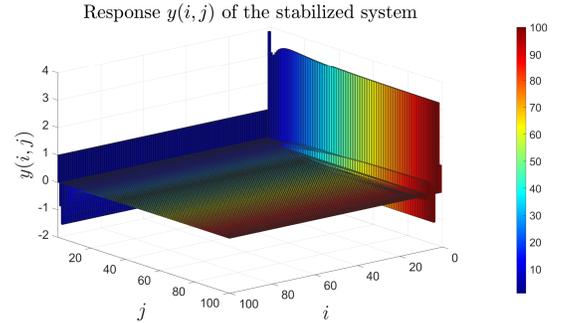

Fig. 4. The output of system (30) after stabilization by output-feedback controller.

Referring to Fig. 2, the zero-input response of the system is observed to tend toward infinity, thus leading to the conclusion that Theorem 1 is validated, indicating system instability.

Subsequently, upon implementing the state-feedback controller derived from Theorem (2), the closed-loop system attains stability. The state-feedback controller matrix K was obtained as $K_{state-feedback} = [-10, 21.0084]$, following the solution of LMIs (15). The complete response of the system (30) with the installed state-feedback controller is illustrated in Fig. 3. It is evident that with the increase in $i$ and $j$, the system's response tends towards zero, indicating stability. $i$ and $j$ respectively describe the spatial and time characteristics of heat in (25), while $y(i, j)$ defines the heat level in terms of time and location. Therefore, the concept of stability in Fig. 3. entails that the temperature level remains bounded for all times and locations.

Fig. 4. illustrates the stability of the closed-loop system response for the output feedback controller with the matrix $K_{\text{output-feedback}} = [-10 \quad 21.0084]$. Based on Fig. 4., it can be observed that over time and space, the system's response has stabilized and tends towards zero.

Research on the case study described by (25) can be extended further to obtain additional numerical results. Depending on the physical conditions, the system defined in (25) can be explored using different discretization scales $\Delta x$ and $\Delta t$, as well as various initial conditions for $x_h(0,0)$ and $x_v(0,0)$.

TABLE I
NUMERICAL RESULTS

| Type of Controller | Matrix K | $\Delta x$ | $\Delta t$ |
|---|---|---|---|
| State-feedback | $[-10, 21.0084]$ | 0.1 | 0.1 |
| Output-feedback | $[-0.8137, 0.5039]$ | 0.1 | 0.1 |

Table I presents a summary of the numerical results obtained in Section IV. These results are only valid for the discretization scales $\Delta x = 0.1$ and $\Delta t = 0.1$. In other words, the system (25) exhibited instability at the discretization scale of $\Delta x = \Delta t = 0.1$ for which the controllers were specifically designed to achieve stabilization.

## V. CONCLUSION

This paper centers on analyzing the stability and achieving of state- and output-feedback controllers for two-dimensional DAEs described by the Rosser model. The methods proposed in this article introduce a new criterion for stability testing without the necessity of decomposing the system into differential and algebraic components. Furthermore, the paper delves into the achievement and development of state- and output-feedback controllers, obviating the necessity for differential-algebraic decomposition. The efficacy of the proposed methodologies is substantiated via an illustrative industrial case study.

Dynamic controller design can be considered as a future work feature. Additionally, the design of robust or adaptive controllers for n-dimensional DAEs in the presence of uncertainties, disturbances, and delays can also be proposed.

REFERENCES

[1] Cheng S, Paley DA. Cooperative estimation and control of a diffusion-based spatiotemporal process using mobile sensors and actuators. Autonomous Robots. 2023 May 24:1-7.
[2] Brady R, Enderling H. Mathematical models of cancer: when to predict novel therapies, and when not to. Bulletin of mathematical biology. 2019 Oct;81:3722-31.
[3] Kavallaris NI, Suzuki T. Non-local partial differential equations for engineering and biology. Mathematical Modeling and Analysis. 2018;31.
[4] Benzaouia A, Hmamed A, Tadeo F. Two-dimensional systems. Studies in Systems Decision and Control. 2016;28.
[5] Cai C, Wang W, Zou Y. A note on the internal stability for 2-D singular discrete systems. Multidimensional Systems and Signal Processing. 2004 Apr;15:197-204.
[6] Azarakhsh K, Shafiee M. A New 1-D Model for Singular 2-D Systems. In2021 29th Iranian Conference on Electrical Engineering (ICEE) 2021 May 18 (pp. 607-612). IEEE.
[7] Luenberger D. Dynamic equations in descriptor form. IEEE Transactions on Automatic Control. 1977 Jun;22(3):312-21.
[8] Chen SF. Stability analysis and stabilization of 2-D singular Roesser models. Applied Mathematics and Computation. 2015 Jan 1;250:779-91.
[9] Linh VH, Ha P. Index reduction for second order singular systems of difference equations. Linear Algebra and its Applications. 2021 Jan 1;608:107-32.
[10] Raj P, Pal D. Lie algebraic criteria for stability of switched systems of differential algebraic equations (DAEs). IEEE Control Systems Letters. 2020 Nov 6;5(4):1333-8.
[11] Atrianfar H. Sampled-time containment control of high-order continuous-time MASs under heterogenuous time-varying delays and switching topologies: A scrambling matrix approach. Neurocomputing. 2020 Jun 28;395:24-38.
[12] Wei Y, Zhao L, Lu J, Alsaadi FE, Cao J. LMI Stability Condition for Delta Fractional Order Systems With Region Approximation. IEEE Transactions on Circuits and Systems I: Regular Papers. 2023 Jun 8.
[13] Malik SH, Tufail M, Rehan M, Rashid HU. Overflow oscillations-free realization of discrete-time 2D Roesser models under quantization and overflow constraints. Asian Journal of Control. 2022 May;24(3):1416-25.
[14] Badie K, Alfidi M, Tadeo F, Chalh Z. Robust state feedback for uncertain 2-D discrete switched systems in the Roesser model. Journal of Control and Decision. 2021 Jul 3;8(3):331-42.
[15] Zamani M, Zamani I, Shafiee M. Robust Control of Positive 2-Dimensional Systems with Bounded Realness Property. Differential Equations and Dynamical Systems. 2022 Sep 27:1-22.
[16] Badie K, Alfidi M, Chalh Z. Further results on $H_\infty$ filtering for uncertain 2-D discrete systems. Multidimensional Systems and Signal Processing. 2020 Oct;31(4):1469-90.
[17] Zhu Z, Lu JG, Zhang QH. Finite Frequency $H_\infty$ Control of Fractional-order Continuous-discrete 2D Roesser Models. IEEE Transactions on Circuits and Systems II: Express Briefs. 2023 Apr 5.
[18] Ouhib L, Kara R. Proportional Observer design based on D-stability and Finsler's Lemma for Takagi-Sugeno systems. Fuzzy Sets and Systems. 2023 Jan 10;452:61-90.
[19] Zamani M, Shafiee M, Zamani I. Stability analysis of singular 2-D positive systems. In2021 29th Iranian Conference on Electrical Engineering (ICEE) 2021 May 18 (pp. 626-631). IEEE.
[20] Chelliq EH, Alfidi M, Chalh Z. Admissibility and robust $H_\infty$ controller design for uncertain 2D singular continuous systems with interval time-varying delays. International Journal of Systems Science. 2023 Apr 26;54(6):1377-97.
[21] Zhang X, Zhang L, Zhao X, Zhao N. Reachable set control for singular systems with disturbance via sliding mode control. Journal of the Franklin Institute. 2023 Mar 1;360(4):3307-29.